\documentclass[global,twocolumn]{svjour}
\newcommand{\mypicwidth}{0.475\textwidth} 

\usepackage{graphics}
\journalname{}
\def\tm{\leavevmode\hbox{$\rm {}^{TM}$}}
\newcommand{\brightstate}{$|\!\!\downarrow\rangle$ }
\newcommand{\darkstate}{$|\!\!\uparrow\rangle$ }

\newcommand{\mbrightstate}{|\!\!\downarrow\rangle}
\newcommand{\mdarkstate}{|\!\!\uparrow\rangle}

\newcommand{\rfig}{Figure }
\newcommand{\rsec}{Section }

\newcommand{\pstate}[1]{$^2$P$_{#1/2}$}
\newcommand{\sstate}{$^2$S$_{1/2}$ }

\newcommand{\mg}{$^{25}$Mg$^+$ }

\newcommand{\mgatom}{$^{25}$Mg }
\newcommand{\mgVatom}{$^{24}$Mg }
\newcommand{\mgSatom}{$^{26}$Mg }

\newcommand{\first}{1st }
\newcommand{\second}{2nd }

\graphicspath{{pics/}}

\begin{document}

\title{A Single Laser System for Ground-State Cooling of \mg}


\author{
Boerge Hemmerling\inst{1}
\and
Florian Gebert\inst{1}
\and
Yong Wan\inst{1}
\and
Daniel Nigg\inst{1,3}
\and
Ivan V.~Sherstov\inst{1}
\and
Piet O.~Schmidt\inst{1,2}
}                     


\institute{QUEST Institute for Experimental Quantum Metrology, Physikalisch-Technische Bundesanstalt, Bundesallee 100, 38116 Braunschweig, Germany\\
\email{boerge.hemmerling@quantummetrology.de} 
\and Institut f\"ur Quantenoptik, Leibniz Universit\"at Hannover, Welfengarten 1, 30167 Hannover, Germany
\and \emph{Present address:} Institut f\"ur Experimentalphysik, Universit\"at Innsbruck, Technikerstrasse 25/4, 6020 Innsbruck, Austria}


\maketitle


\begin{abstract}
We present a single solid-state laser system to cool, coherently
manipulate and detect \mg ions. Coherent manipulation is accomplished
by coupling two hyperfine ground state levels using a pair of far-detuned
Raman laser beams. Resonant light for Doppler cooling and detection is
derived from the same laser source by means of an electro-optic
modulator, generating a sideband which is resonant with the atomic
transition. We demonstrate ground-state cooling of one of the
vibrational modes of the ion in the trap using resolved-sideband
cooling. The cooling performance is studied and discussed by observing
the temporal evolution of Raman-stimulated sideband transitions. The
setup is a major simplification over existing state-of-the-art systems,
typically involving up to three separate laser sources.
\end{abstract}


\section{Introduction}
\label{intro}

Over the last decades, ion trap experiments have become versatile instruments to investigate a variety of physical phenomena, ranging from quantum computing \cite{Blatt:08,Cirac:95}, optical clocks \cite{Chou:10,Rosenband:08} and precision spectroscopy \cite{Schmidt:05,Roos:06,Wolf:09} to quantum simulations \cite{Friedenauer:08,Gerritsma:10,Kim:10}. Many applications benefit from the fact that ions can be easily stored \cite{Paul:90}, with trap life times spanning from hours to months, and coherently manipulated with commercially available laser systems. Furthermore,\\ trapped ions feature very long coherence times \cite{Haeffner:05,Langer:05} and the possibility of cooling the external degrees of freedom to the vibrational ground state \cite{Schaetz:07,Eschner:03,Leibfried:03,Monroe:95,Epstein:07}, making them an ideal test bed to investigate quantum mechanics. The strong Coulomb interaction can then be used to cool simultaneously stored charged particles, opening new prospects for cooling of complex molecular ions to the rotational and vibrational ground state \cite{Schiller:10,Drewsen:10}.

In many experiments, ions with hyperfine structure (e.g. $^{43}$Ca$^+$\cite{Kirchmair:09}, $^9$Be$^+$\cite{Wineland:98}, $^{111}$Cd$^+$\cite{Lee:03}, $^{25}$Mg$^+$\cite{Friedenauer:08}, Yb$^+$ \cite{Olmschenk:07}) are employed. For quantum computing experiments the long-lived hyperfine ground states are usually chosen as qubit states. Ground-state cooling is achieved by driving optical Raman transitions between these states, which often requires several complex Raman and repumping laser systems \cite{Schaetz:07,Wineland:98}. Here, we report on cooling of the \mg isotope to an average vibrational population of $\bar{n} = 0.03 \pm 0.01$ using a single solid-state fiber laser system with an incorporated electro-optical modulator. The principle behind the laser setup represents a major simplification over previously used Mg$^+$ laser systems \cite{Friedenauer:06}. A similar approach based on acousto-optical modulators with less detuning of the Raman beams has yielded a mean vibrational population of $\bar{n} = 0.34 \pm 0.08$ \cite{Epstein:07}.

This paper is organized as follows. In Section \ref{setup} the experimental setup used for trapping, cooling and manipulating a single \mg ion in a Paul trap is described. In Section \ref{gsc_cooling} the sequences to achieve ground-state cooling of the \mg ion are explained and in Section \ref{results} the experimental results on the cooling efficiency are discussed.

\section{Setup}
\label{setup}

The following sections contain a detailed description of our experimental setup to implement Doppler and Raman resolved-sideband cooling of an \mg ion stored in a linear Paul trap.

\subsection{Solid-State Laser System for Magnesium}
\label{setup:laser_system}

\begin{figure}
\centerline{\resizebox{\mypicwidth}{!}{\includegraphics{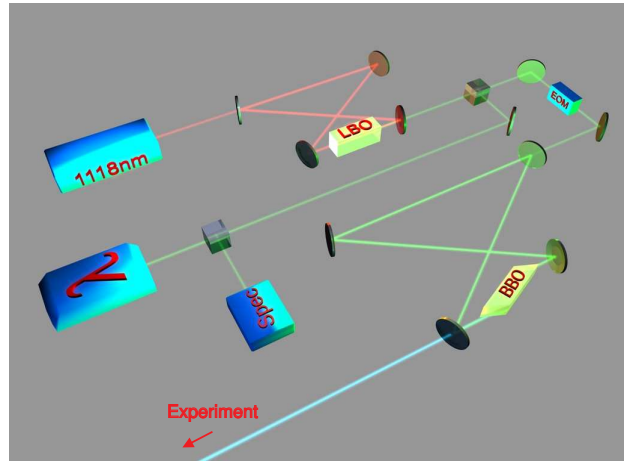}}}
\caption{Optical setup of the magnesium laser system. The output of a fiber laser at $1118$\,nm is frequency-quadrupled in two SHG cavities to drive transitions in \mg ions (see text for details). Typically $15$\,mW of the green light is redirected to a saturated iodine spectroscopy (Spec) for frequency-locking of the fiber laser and a wavelength-meter for monitoring. An EOM imprints sidebands at $9.2$\,GHz on the green laser and allows for fast switching between a resonant Doppler cooling and an off-resonant Raman configuration.}
\label{fig:laser_setup}       
\end{figure}

The Doppler cooling transition in \mg is at a wavelength of 280\,nm. We generate this light by frequency quadrupling a commercial fiber laser\footnote{Koheras Boostik\tm Y10/Menlo Systems GmbH orange one-1} at $1118.21$\,nm with a specified line width of $<\!70$\,kHz and an output power of $1$\,W, similarly to the setup presented in \cite{Friedenauer:06}. As shown in \rfig\ref{fig:laser_setup}, the output of the laser is frequency-doubled in a second-harmonic generation (SHG) bow-tie cavity by an anti-reflection-coated $4 \times 4 \times 18$\,mm$^3$ Lithium Triborate (LBO) crystal\footnote{Castech Crystals Inc.} using $90^\circ$ non-critical phase-matching of type I. A typical output power of $\sim\!450$\,mW at $559$\,nm is observed after the cavity. Usually $15$\,mW of the light is picked up for frequency-stabilizing the fiber laser using saturation spectroscopy in a molecular $^{129}I_2$ cell.

The output of the LBO cavity is coupled into a second bow-tie cavity which contains a $3 \times 3 \times 10$\,mm$^3$ brewster-cut $\beta$-Barium-Borate (BBO) crystal\footnote{D\"ohrer Elektrooptik} for SHG using critical phase-matching of type I. Here, a typical output power of $\sim\!60$\,mW at $280$\,nm is observed. The length of each cavity is stabilized by the standard H\"ansch-Couillaud locking technique \cite{Haensch:80}.

The novelty in this setup consists of an electro-optical modulator\footnote{Laser 2000 GmbH NFO-4851-M} (EOM) that is located between the two doubling cavities to imprint sidebands at $9.2$\,GHz onto the green laser beam. The fiber laser is frequency-tuned such that one of the sidebands coincides with the \sstate to \pstate{3} transition of $^{25}\textrm{Mg}^+$, which is used for Doppler cooling. The optical carrier, detuned by $9.2$\,GHz when sidebands are switched off, is employed for driving Raman transitions between two hyperfine states, as shown in \rfig\ref{fig:mg25_level_scheme} and described in \rsec\ref{setup:coherent_manipulation}. This setup allows for fast switching between the resonant Doppler cooling and the off-resonant Raman configuration on a timescale faster than $\mu$s while using just a single laser system.

It is worthwhile mentioning that the frequency doubling process affects the modulation index, but does not change the carrier-sideband frequency separation \cite{Lee:03}. The EOM phase-modulates the electric field with frequency $\omega$ as
\begin{equation} 
e^{i\omega t} \longrightarrow e^{i (\omega t + \beta \sin \Omega t) },
\end{equation}
where $\beta$ is the modulation index and $\Omega = 2\pi \times 9.2$\,GHz in our case. Frequency doubling then doubles the modulation index, since
\begin{eqnarray}
\nonumber
e^{i (\omega t + \beta \sin \Omega t) } \quad {\textrm{SHG} \atop \longrightarrow} \quad e^{i (2\omega t + 2\beta \sin \Omega t) }.
\end{eqnarray}
The transmission of both $2\omega \pm \Omega$ sidebands through the BBO cavity is guaranteed by adjusting its free spectral range accordingly. We typically drive the EOM with $0.7$\,W of radio-frequency power to achieve a measured carrier to single-sideband ratio of $\sim 11$ before frequency doubling, which corresponds to a modulation index of $\beta\sim 0.58$. After frequency doubling, we measure $25\%$ of the total power in each of the sidebands as expected for twice the modulation index. It should be noted that driving the EOM with more than $1.3$\,W results in significant beam deflection when turning the EOM on. This is probably caused by thermal effects in the EOM crystal.

\begin{figure}
\centerline{\resizebox{\mypicwidth}{!}{\includegraphics{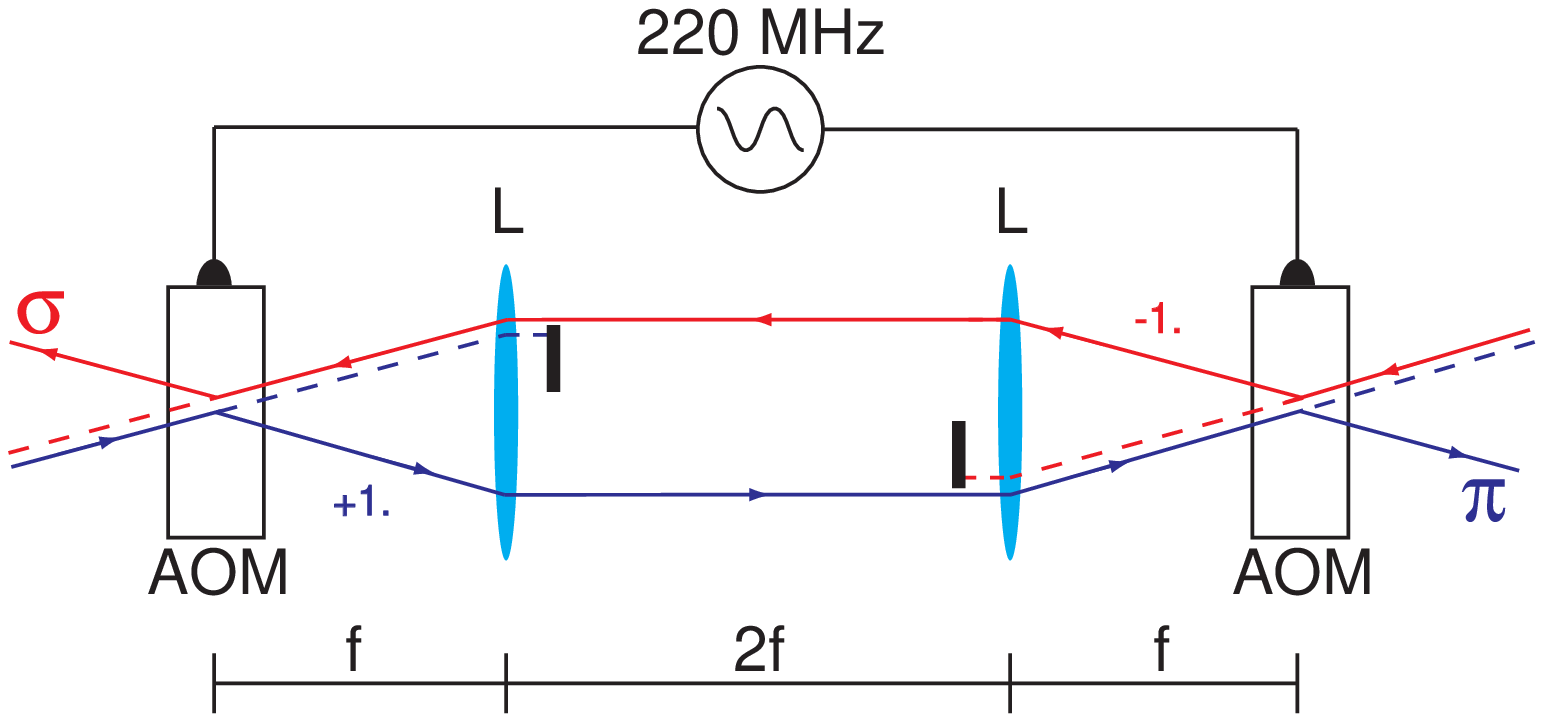}}}
\caption{Schematic of the double-pass AOM setup. The $\sigma$- and $\pi$-Raman beams are produced by two beams that counter-propagate. The 0th order (dashed lines) of each AOM is blocked. The use of the lenses (L) allows for a frequency scan without spatially shifting the outgoing beams. Both AOMs have to be supplied by the same frequency to fulfill the Bragg condition for each beam. The incoming and the outgoing overlapping beams are separated by translating them by a few millimeters (indicated by the small shift). This configuration avoids foci near optical surfaces.}
\label{fig:aom_setup}       
\end{figure}

For the Raman configuration the laser beam is split into two branches. Each branch passes through an individual single-pass AOM which is resonantly driven at $450$\,MHz\footnote{Brimrose Corp. QZF-450-100-.280}. After that, the beams counter-propagate through the same AOM double-pass configuration (see \rfig\ref{fig:aom_setup}) consisting of two AOMs, driven at $\sim\!220$\,MHz\footnote{IntraAction Corp. ASM-2202B3}, providing two Raman laser beams ($\sigma$ and $\pi$) with a frequency difference of $1.789$\,GHz. Typically, we achieve an efficiency of $\sim\!70$\% for the single-pass AOMs and $\sim\!50$\% for the double-pass AOMs for each branch.

This configuration, in contrast to otherwise used double-pass setups \cite{Chang:08,Donley:05}, has been chosen for two reasons. Firstly, the most common way to realize a double-pass AOM is by turning the linear polarization after the first pass of the AOM by $90^\circ$ and then separating the overlapped retro-reflected beam from the incoming beam with a polarizing beam splitter . This is not possible in the ultra-violet (UV) since the diffraction efficiency of the UV grade fused silica crystal of the AOM strongly depends on the polarization. In a second approach, the beams are geometrically separated. This is usually done by a combination of a lens and a cat's eye which can for example be realized by two reflecting mirrors or a single right-angle prism. Using both these configurations in our setup, we have observed a degradation of the laser beam shape along with a significant decrease of the overall light transmission efficiency over a timescale of a few hours. We attribute this effect to UV-induced damage on the optical cathetus facet of the prism and on the mirror surfaces, respectively. Since a degradation was only observed on the first reflecting surfaces that were hit by the laser beam, we assume that the light intensity has to overcome a certain threshold to produce this effect \cite{Negres:10}. In both cases, typically a $\sim\!5$\,mW laser beam was focused to a waist of $50 - 70\,\mu$m at the prism or mirror, resulting in an intensity of $320 - 640$\,kW/m$^2$. In order to circumvent this effect, we re-designed the double-pass AOM configuration, according to \rfig\ref{fig:aom_setup}, avoiding a focus near any of the optical surfaces.

The AOM setup provides the laser beams for off-res\-onant Raman excitation if the EOM is switched off. At the same time, the $\sigma$-Raman beam can be used to resonantly repump the population from the upper hyperfine manifold of the ground state (F$=\!2$) by switching on the EOM. This demonstrates the simplicity of this setup, since otherwise a second laser system providing an individual resonant repumping beam is necessary.

Our experimental sequences require pulsed operation, hence all laser beams are intensity stabilized by means of a sample-and-hold proportional-integral controller. A field-programmable gate array (FPGA) based pulse sequencer, which allows for frequency and phase-switch\-ing of six direct digital synthesis (DDS) output channels at a sub-$\mu$s rate \cite{Pham:05}, is used to provide frequency sources for all AOMs. Every time-critical component in the setup is referenced during all the measurements discussed here to a $10$\,MHz frequency standard\footnote{Stanford Research Systems DS-345}.

\subsection{Ion Trap}
\label{setup:ion_trap}

We produce \mg ions using photoionization of a thermal beam of magnesium atoms provided by resistive heating of a stainless-steel tube oven. A single light source at $285$\,nm with a power of $\sim\!300$ $\mu$W from a frequency-quadrupled diode laser is used for this process\footnote{Toptica DL pro}. The diode laser has an output power of $100$\,mW and operates at a wavelength of $1140$\,nm. Its output is frequency-quadrupled by means of two cascaded SHG cavities, similarly to the main laser setup\footnote{In this case, the cavities contain a LiNbO$_3$ $\cdot$ MgO doped crystal (HG Photonics) and a BBO crystal (Castech Crystals Inc.) for SHG. Both cavities are stabilized using the Pound-Drever Hall locking technique \cite{Drever:83}.}.

The atoms are ionized in a two-step process, involving the $^1P_1$ state in neutral magnesium \cite{Kjaergaard:00}. Although the natural abundance of the \mgatom isotope is only $\sim\!10$\%, it turns out to be sufficient for satisfactory loading rates of the desired isotope. The setup allows for isotope-selective loading by tuning the frequencies of the ionization and Doppler cooling lasers according to the isotope shifts. The frequency of the \mgatom (\mgSatom\hspace{-0.1cm}) resonance is higher by $\sim 744$\,MHz ($\sim 1414$\,MHz) compared to the \mgVatom resonance, respectively \cite{Salumbides:06}.

\begin{figure}
\centerline{\resizebox{\mypicwidth}{!}{\includegraphics{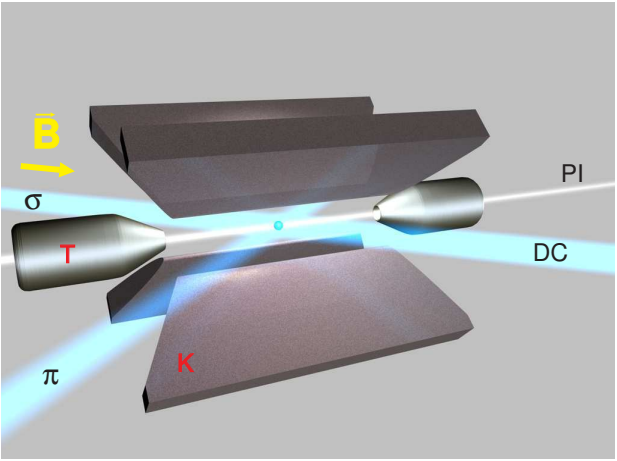}}}
\caption{Linear Paul trap. The trap consists of four blades (K) for radial confinement and two tip electrodes (T) for axial confinement. One pair of opposing blades is connected to a radio-frequency source at $23.81$\,MHz and the other pair is grounded. A DC voltage of $2$\,kV is applied to the tip electrodes. The distance between opposite blades is $1.6$\,mm, and between the tips it is $5$\,mm. The photoionization laser at $285$\,nm (PI) is applied onto the atom through the tip electrodes. The Raman lasers ($\sigma$, $\pi$) are used in a crossed-configuration at an angle of $45^\circ$ with respect to the trap axis to provide an effective wave vector along the trap axis. The Doppler cooling beam (DC) is counter-propagating with the $\sigma$ beam. A magnetic field of $\vec{B}\!\sim\!0.6$ mT is co-aligned with the DC beam.}
\label{fig:trap}       
\end{figure}

After the ionization process, a single \mg ion is stored in a linear Paul trap \cite{Gulde:03}, as shown in \rfig\ref{fig:trap}. Axial confinement is provided by applying a DC voltage of typically $2$\,kV to the tip electrodes and radial confinement is realized by applying a radio-frequency of $7$\,W at $23.81$\,MHz, which is then resonantly enhanced by means of a helical resonator \cite{Macalpine:59}, yielding an amplitude on the order of a few hundreds of volts at the radial blades. These parameters result in axial and radial harmonic confinement with trapping frequencies for \mg of $(\omega_{\textrm{\tiny ax}},\omega_{\textrm{\tiny rad}}) \sim 2\pi \times (1.9,2.3)$\,MHz.

\subsection{State Initialization and Coherent Manipulation}
\label{setup:coherent_manipulation}

\begin{figure}
\centerline{\resizebox{\mypicwidth}{!}{\includegraphics{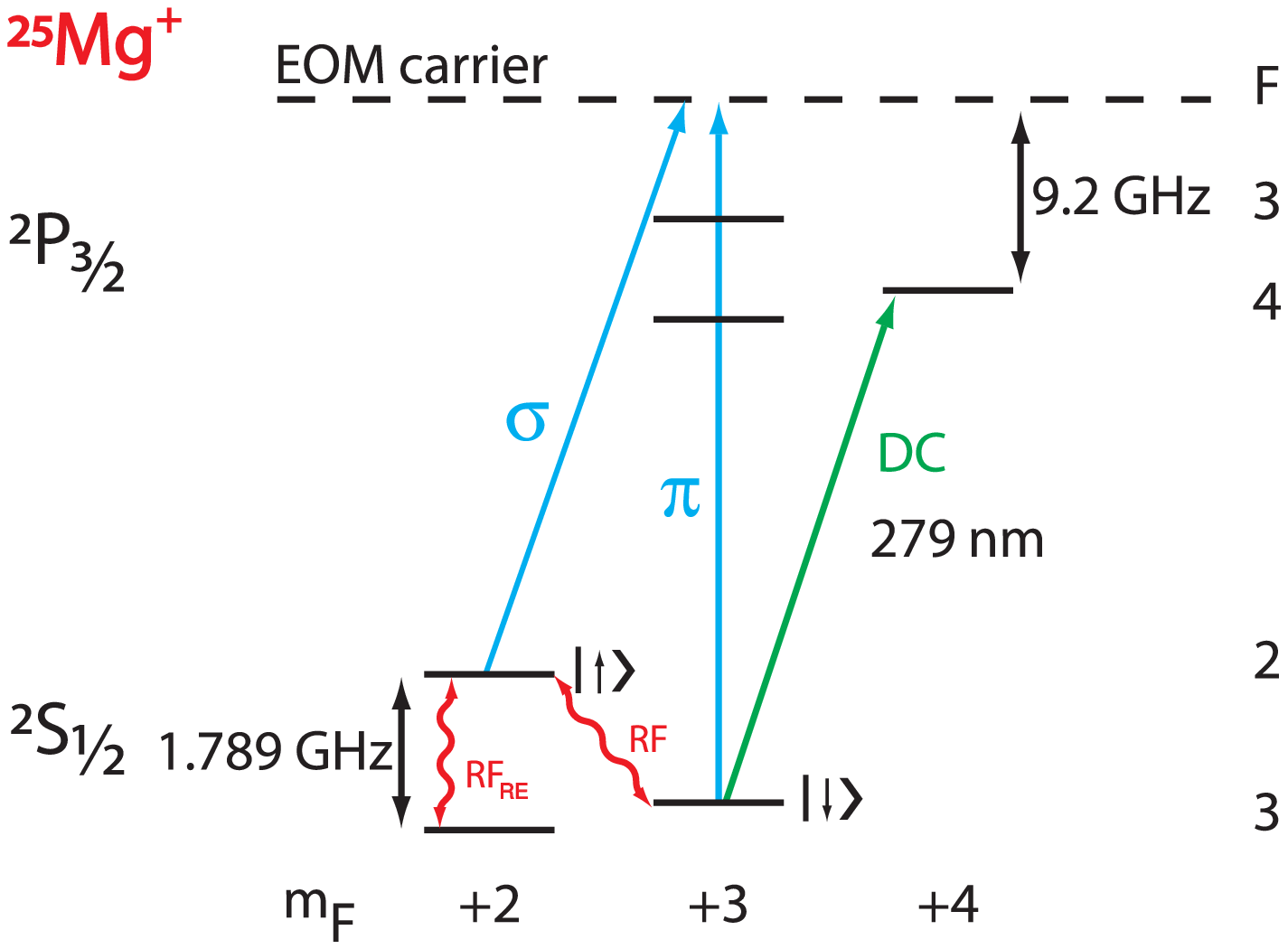}}}
\caption{Partial \mg Level Scheme. The laser is tuned $9.2$\,GHz off-resonant with respect to the \sstate to \pstate{3} transition. The optical carrier is used to drive coherent Raman transitions between the $\mbrightstate := \left|\textrm{F}=3,\textrm{m$_\textrm{\tiny F}$}=3\right\rangle$ and $\mdarkstate := \left|2,2\right\rangle$ states by the $\sigma$- and $\pi$-Raman beams. One of the optical sidebands induced by an EOM is employed for resonant Doppler cooling on the cycling transition between the \sstate $\left|3,3\right\rangle$ and \pstate{3} $\left|4,4\right\rangle$ states (DC beam). Magnetic field induced hyperfine transitions are driven by applying radio-frequency pulses at $1.789$\,GHz (RF and RF$_{\textrm{\tiny RE}}$).}
\label{fig:mg25_level_scheme}       
\end{figure}

The transitions used for cooling and coherent manipulation of the states of the ion are shown in the simplified level scheme of \mg in \rfig\ref{fig:mg25_level_scheme}. During all measurements the degeneracy between the magnetic sub-levels is lifted by a magnetic field of $\vec{B}\!\sim\!0.6$\,mT. Doppler cooling of the ion is achieved by switching on the EOM sidebands, providing a near-resonant coupling of one of the sidebands to the \sstate to \pstate{3} transition, which has a natural line width of $\gamma = 2\pi \times 41.4 $\,MHz \cite{Nist}.
 
To perform optimal cooling, the beam is detuned by half the line width by means of acousto-optical modulators (AOM) and its power is set to the saturation intensity of the transition ($\sim 250$\,mW/cm$^2$). The beam's polarization is carefully adjusted such that only $\sigma^-$ ($\Delta m_\textrm{\tiny F} = +1$) transitions are addressed. While applying this beam, the ion is optically pumped into the \brightstate state, ideally leading to a closed cycling transition. Population in the F$=\!2$ manifold is efficiently transferred to the F$=\!3$ manifold by a repumping beam (see below) which is additionally applied during Doppler cooling to further improve state initialization. 
 
For coherent manipulation of the ion, we couple the hyperfine ground states, \brightstate and \darkstate\hspace{-0.1cm}, either by a Raman transition which is off-resonant by $9.2$\,GHz to the \sstate to \pstate{3} transition or by direct application of radio-frequency at $1.789$\,GHz, corresponding to the hyperfine splitting in \mg \cite{Itano:81}.
For the given Rabi frequencies in this work, each Raman beam has typically $\sim 1$\,mW of power in a collimated beam of $\sim 500\,\mu$m diameter before being focused onto the ion by a lens (f=$200$\,mm). In case of the radio-frequency transitions, sufficiently strong field amplitudes at the position of the ion are achieved by applying $\sim 4$\,W of microwave power to an impedance-matched quarter-wave antenna made of a special coaxial cable\footnote{Andrew CommScope FSJ1-50A} with an attenuation of $0.27$\,dB/m at $1.8$\,GHz. The antenna is attached to the outside of the vacuum chamber (see \rfig\ref{fig:chamber}) at a distance of $\sim\!12$\,cm from the ion.

\subsection{Fluorescence and State Detection}
\label{setup:detection}

\begin{figure}
\centerline{\resizebox{\mypicwidth}{!}{\includegraphics{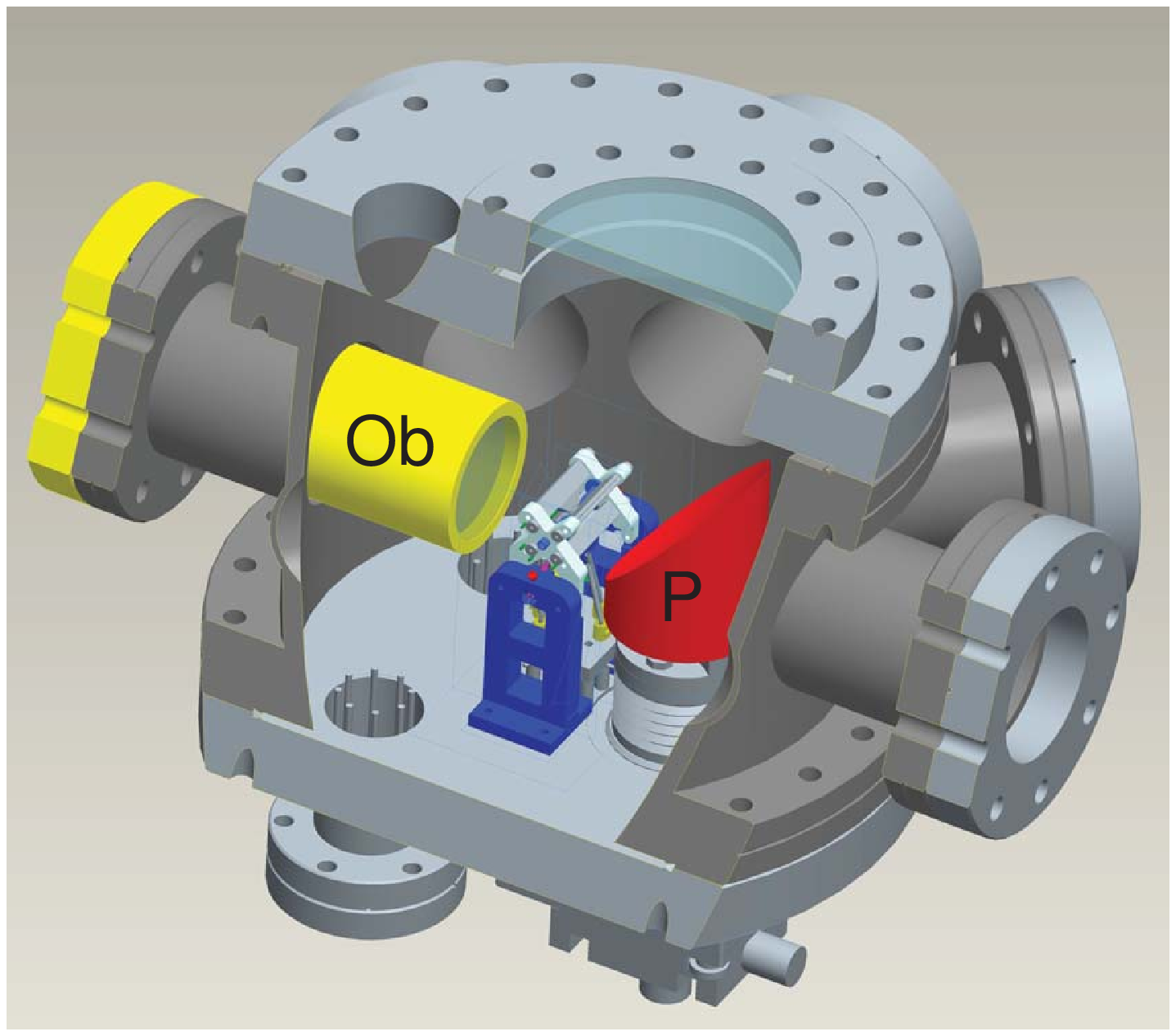}}}
\caption{Schematic cross section of the vacuum setup. Fluorescence from the ion stored in the Paul trap (center) is collected via an UV objective (Ob) and a parabolic mirror (P). The objective is mounted inside an inverted viewport to minimize the distance to the ion for higher collection efficiency. Both channels are imaged outside the vacuum chamber onto a single PMT, yielding their combined count rates.}
\label{fig:chamber}       
\end{figure}

The fluorescence of the ion resulting from the excitation of the resonant Doppler cooling beam is collected by both an UV objective and an aluminum parabolic mirror (see \rfig\ref{fig:chamber}) and imaged onto a single photon counting module\footnote{Hamamatsu Photonics H8259 MOD with R7518P} (PMT) for quantitative readout. The fluorescence light can also be directed onto a CCD camera\footnote{Andor Technology iXon$^{\textrm{EM}}$+ 885 EMCCD} to observe a geometrical image of the ion for alignment purposes. The UV objective is an extension of an existing design \cite{Alt:02} covering a simulated solid angle of $\sim\!3\!-\!4$\% of $4\pi$, whereas the parabolic mirror\footnote{Kugler Precision GmbH, custom-made} covers a solid angle of $\sim\!11$\% for fluorescence collection \cite{inprep}. Typically, a count rate of $\sim\! 240$\,kHz is observed from a single \mg ion during Doppler cooling, when the laser is red-detuned by $\sim\! 20$\,MHz from the \sstate to \pstate{3} transition and set to saturation intensity.

In all measurements described below, the ion is first initialized in the \brightstate state by applying the Doppler cooling and repumping laser beams for $\sim\!1$ ms and then coherently manipulated according to the specific experimental sequence of laser pulses. After that, state detection is achieved via electron-shelving by applying the resonant Doppler cooling beam for typically $\sim\!10\!-\!15$ $\mu$s and counting the number of detected photons. In contrast to electron-shelving schemes in other ion species \cite{Leibfried:03,Wineland:98}, the detection pulse is strongly restricted in length since the ion can be off-resonantly depumped from the \darkstate state, thus imitating a bright ion which was initially dark. Given this limitation, we typically detect a mean number of $\sim\!5.8$ photons for a bright ion, whereas we observe a mean number of $\sim\!0.2$ photons if the ion was dark.

\begin{figure}
\centerline{\resizebox{\mypicwidth}{!}{\includegraphics{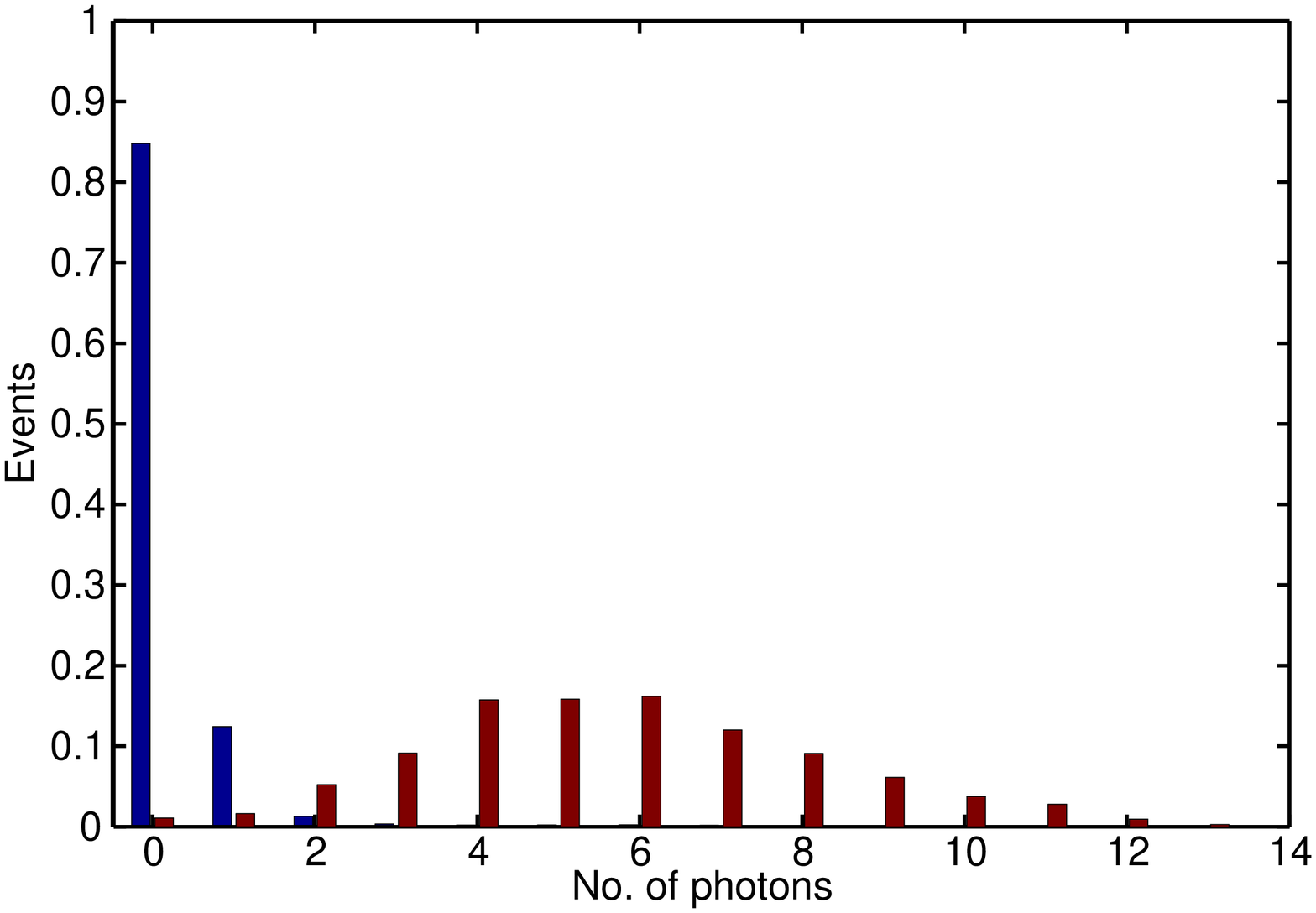}}}
\caption{Dark and bright state histograms. Typical calibrated histograms of the collected number of photons for a dark ion (blue bars) where only stray light is detected, and for a bright ion (red bars), where the ion is Doppler-cooled and prepared in the \brightstate state. The average number of photons in this case is $0.2$ and $5.8$, respectively, for a $12.5$ $\mu$s exposure time.}
\label{fig:psi_dark_and_bright}
\end{figure}

The actual state amplitude $a$ of the ion is determined by repeating the experiment $100 - 300$ times and fitting the resulting photon number distribution to previously calibrated, independent model distributions $\psi = a \psi_\downarrow + (1-a) \psi_\uparrow$, as shown in \rfig\ref{fig:psi_dark_and_bright}. The distribution $\psi_\uparrow$ for a dark ion is recorded by switching off the EOM sidebands during the detection process, thus only accounting for stray light on the PMT, whereas the distribution $\psi_\downarrow$ for a bright ion is taken while the ion is Doppler-cooled and prepared in the \brightstate\hspace{-0.15cm}. Due to the rather low number of detected photons, both distributions are not orthogonal, resulting in an overlap integral of $\sum \psi_\downarrow\psi_\uparrow\sim\!1.4$\%, which limits the single-shot readout fidelity. 

\subsection{Radio-Frequency Rabi Oscillations}
\label{setup:rf_flops}

\begin{figure}
\centerline{\resizebox{\mypicwidth}{!}{\includegraphics{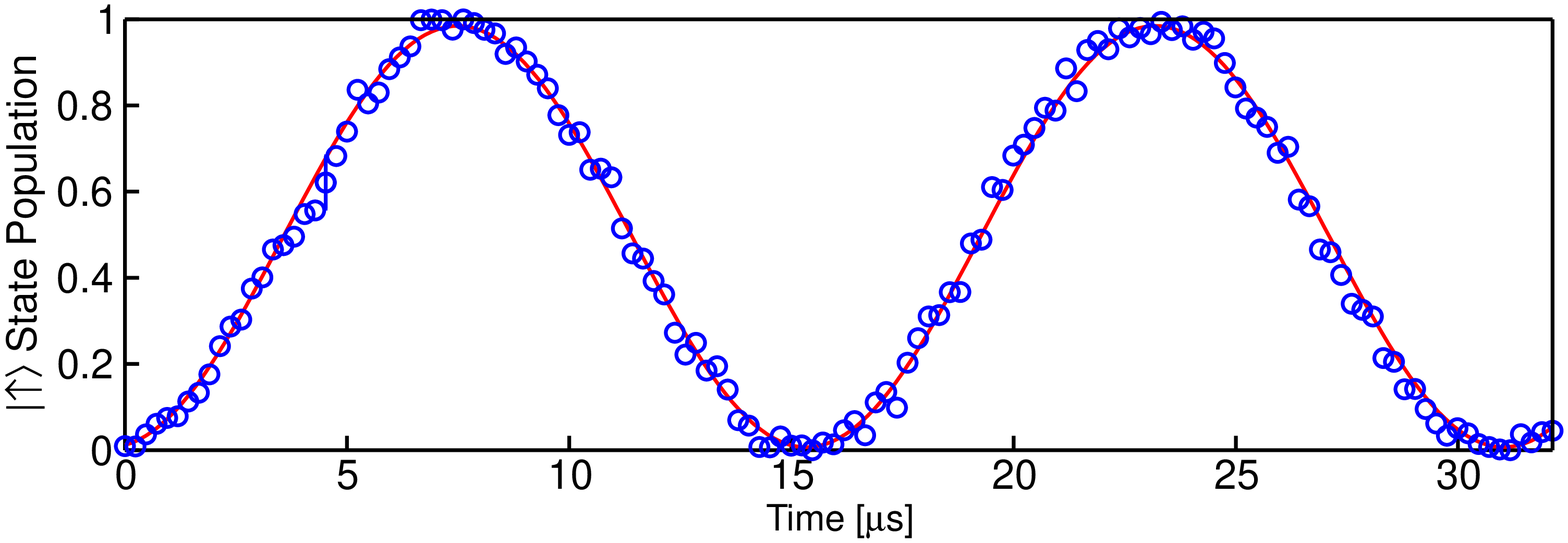}}}
\caption{Radio-frequency induced Rabi oscillations between the \brightstate and \darkstate states. Each point corresponds to $300$ measurements. The average error resulting from the dark and bright state distribution fit is on the order of $\pm 3\%$ for each point and is omitted in the graph for clarity. A sinusoidal fit (solid red line) yields a Rabi frequency of $\Omega_{\textrm{\tiny RF}} = 2\pi \times 63.74(7)$\,kHz and a contrast of $97.8 \pm 1.4$\% for the oscillation.}
\label{fig:rf_rabi_flops}
\end{figure}

As an initial measurement, we observe Rabi oscillations between two hyperfine ground states of \mg \hspace{-0.15cm}. In \rfig\ref{fig:rf_rabi_flops} the probability of finding the Doppler-cooled ion in the \darkstate state after subjecting it to a radio-frequency pulse at $1.789$\,GHz is shown as a function of the pulse duration. Applying the calibrated state detection procedure, we observe the expected sinusoidal state evolution with a Rabi frequency of $\Omega_{\textrm{\tiny RF}} = 2\pi \times 63.74(7)$\,kHz and a contrast of $97.8 \pm 1.4$\%.

It is worth mentioning that the motion of the ion can be neglected in this measurement, since the Lamb-Dicke criterion is fulfilled for the radio-frequency transition \cite{Wineland:98}. 

\section{Ground-State Cooling}
\label{gsc_cooling}

\begin{figure}
\centerline{\resizebox{\mypicwidth}{!}{\includegraphics{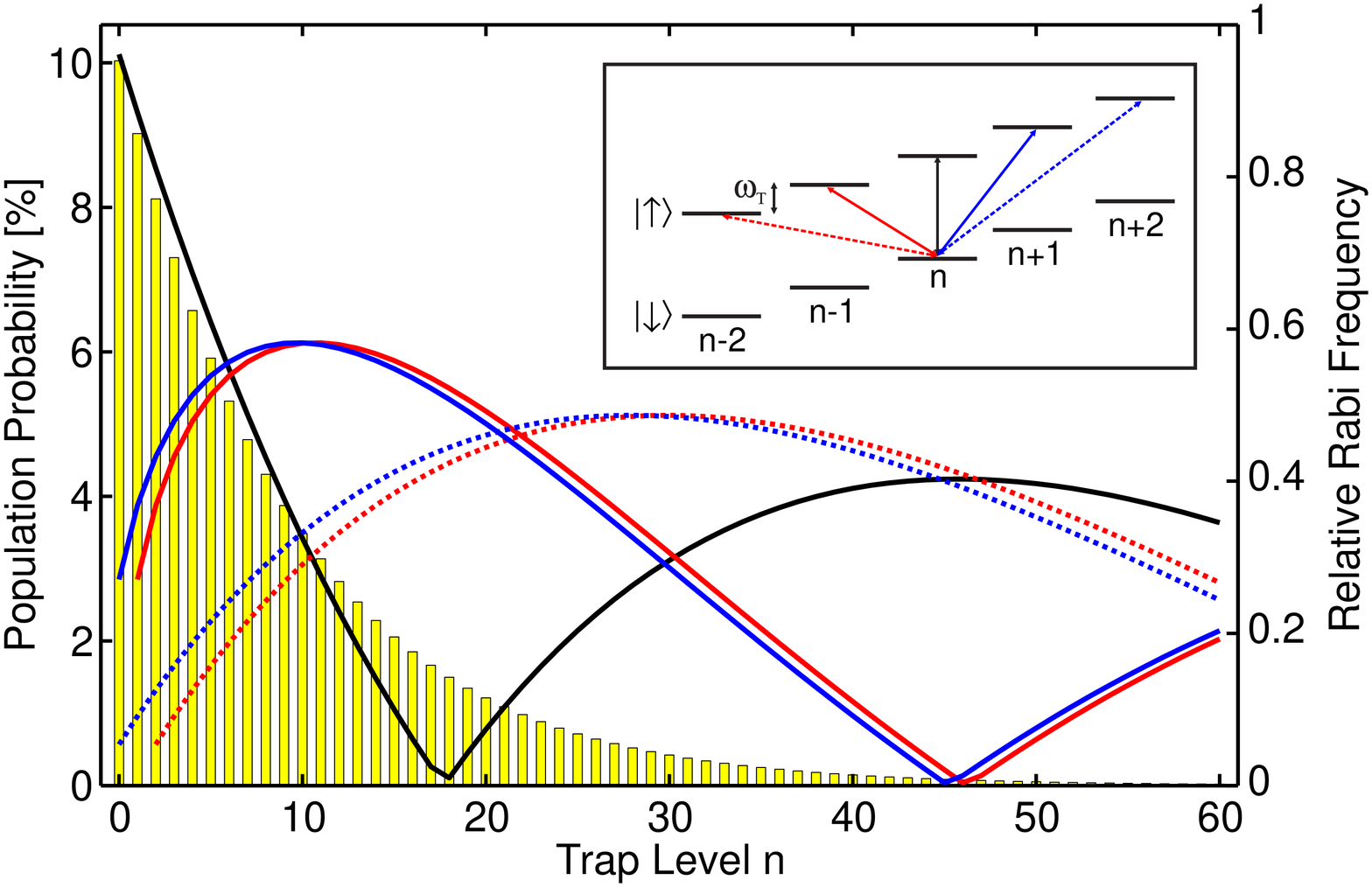}}}
\caption{Population distribution and Rabi frequencies. The bar graph (yellow) shows the population distribution for an \mg at the Doppler cooling limit ($1$\,mK) in a 1D harmonic potential with a trapping frequency of $\omega_T = 2 \pi \times 2.2$\,MHz. In addition, the functional dependence of the Rabi frequencies of the carrier (black line), \first red (blue) sideband (red (blue) solid lines) and \second red (blue) sideband (dashed lines) are shown for a Lamb-Dicke factor of $\eta=0.28$ as a function of the initial $n$-state. The inset schematically depicts the different transitions (same color coding) for the effective two-level system in the harmonic trap.}
\label{fig:rabi_freq}       
\end{figure}

After Doppler cooling, the \mg ion reaches a thermal state with a predicted temperature of $T_{\textrm{\tiny DC}} \sim 1$\,mK \cite{Wineland:79}. This corresponds to a mean vibrational population number of the harmonic trapping potential $\bar{n} \sim 10$, at a trapping frequency of $\omega_T \sim 2\pi \times 2$\,MHz. The predicted population distribution for these parameters is depicted in Figure \ref{fig:rabi_freq}.

We further cool a single ion in axial direction via Raman sideband cooling \cite{Monroe:95} by applying first and second order red sideband (\first\hspace{-0.15cm}/\second RSB) pulses which couple the internal (electronic) with the external (vibrational) degrees of freedom. This is accomplished by detuning both Raman beams by one or two trapping frequencies and adjusting the pulse duration for each target state according to the calculated Rabi frequencies in \rfig\ref{fig:rabi_freq} \cite{Wineland:98}. Reinitialization and dissipation are achieved by applying a resonant repump laser pulse using the $\sigma$-polarized Raman beam with the EOM switched on. From the excited state, the ion can decay into the target state ($\left|\downarrow\right\rangle$), but also into the $\left|3,2\right\rangle$ state. This population is recovered by applying a radio-frequency $\pi$-pulse between the $\left|3,2\right\rangle$ and $\left|2,2\right\rangle$ states (RF$_{\textrm{\tiny RE}}$), followed by another optical repump pulse. Repeating this sequence several times reinitializes the atomic population into the \brightstate\hspace{-0.15cm} state \cite{Chou:10}. Each RSB pulse-repump sequence removes approximately one quantum of motion from the ion in the trap, until the ground state is achieved.

The complete sideband cooling sequence begins with a series of $25\times$ \second RSB pulses, starting from $n_\textrm{\tiny ax}\!\sim\!40$, which are followed by $15\times$ \first RSB pulses, starting from $n_\textrm{\tiny ax}\!\sim\!15$, to cool the ion to the ground state. Each set of pulses is usually repeated $2-3$ times to improve the cooling process. The use of the \second RSB in this procedure is necessary, since ion population in $n$-levels beyond the zero crossing of the Rabi frequency of the \first RSB cannot be efficiently transferred below this point using only the \first RSB. The total sideband cooling sequence typically takes $\sim\!10-15$ ms.

\section{Results}
\label{results}

\begin{figure}
\centerline{\resizebox{\mypicwidth}{!}{\includegraphics{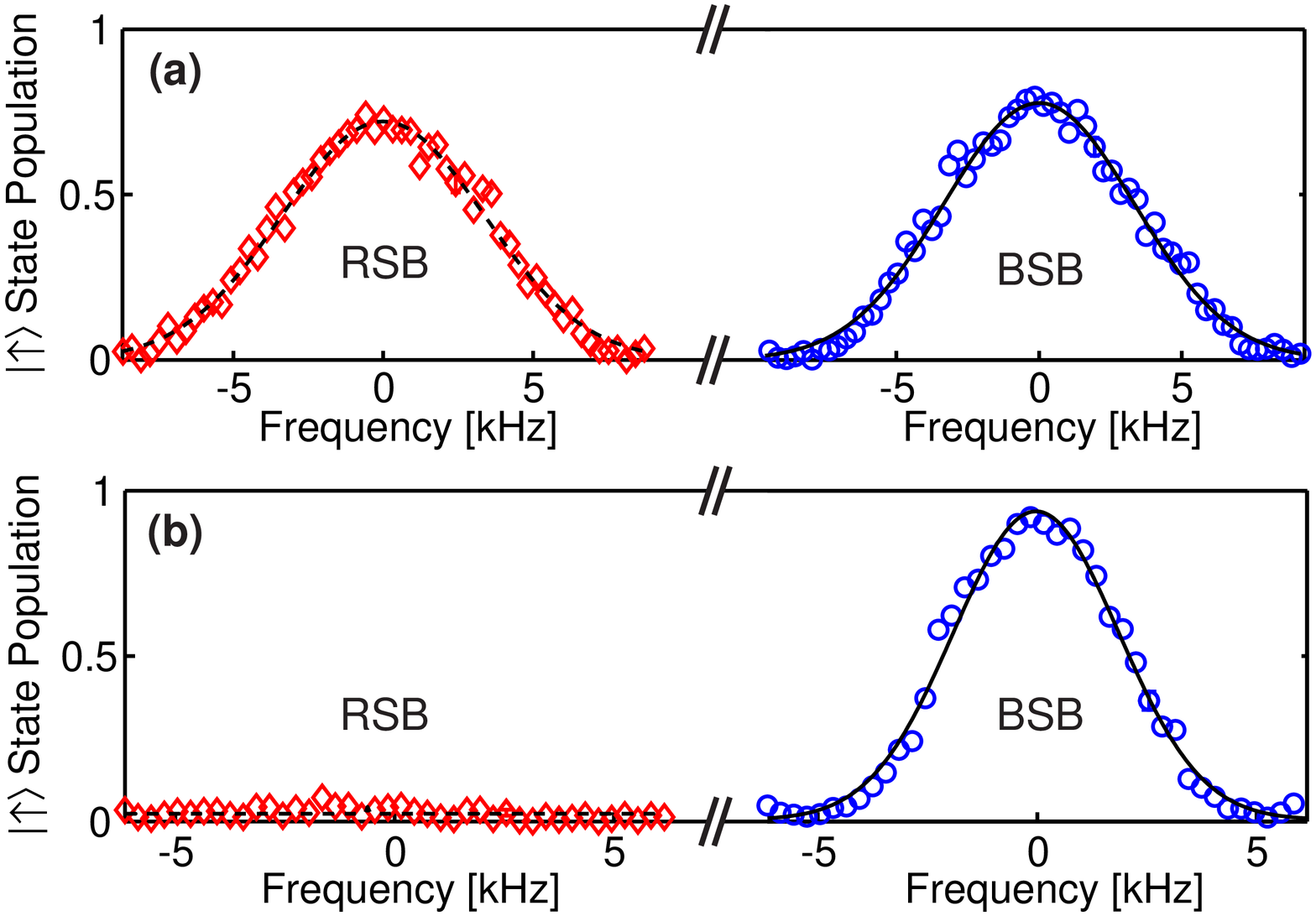}}}
\caption{Sideband excitations. Frequency scans over the \first red and blue sidebands (diamonds and circles) for a Doppler-cooled {\bf (a)} and a sideband-cooled {\bf (b)} \mg ion are shown. The excitation pulse has a length of {\bf (a)} $25 \mu s$ and {\bf (b)} $45 \mu s$, respectively. The missing excitation on the red sideband clearly indicates the high ground state occupation. The frequencies are shifted for clarity. The average error resulting from the dark and bright state distribution fit is on the order of $\pm 3\%$ for each point and is omitted in the graph for clarity. The solid (dashed) lines correspond to Gaussian fits of the red (blue) sideband resonance.}
\label{fig:gsc_sidebands}       
\end{figure}

\rfig\ref{fig:gsc_sidebands} shows the measured excitation on both RSB and BSB resonances for a Doppler-cooled and an axially sideband-cooled single \mg ion, respectively. As the ground state is expected to be a dark state for the red sidebands, the \first RSB shows no excitation after sideband cooling, whereas the \first BSB excitation rises almost to unity. The ratio $Q = \rho_R/\rho_B$ of the RSB ($\rho_R$) and BSB ($\rho_B$) excitations gives the average axial vibrational population $\bar{n}_{\textrm{\tiny ax}}$ \cite{Turchette:00}
\begin{equation}
\bar{n}_{\textrm{\tiny ax}} = \frac{Q}{1-Q},
\end{equation}
as well as the population in the axial ground state $P(n = 0) = 1 - Q$. The analysis of the observed resonances, along with Rabi oscillations on the sidebands, yields $\bar{n}_{\textrm{\tiny ax}} = 0.03 \pm 0.01$ as an upper limit for the mean occupation. This corresponds to a ground state population of $P(n = 0) \sim 97\%$.

\begin{figure}
\centerline{\resizebox{\mypicwidth}{!}{\includegraphics{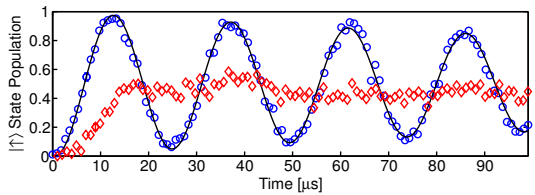}}}
\caption{Raman-stimulated Rabi oscillations on the carrier transition for a Doppler-cooled (diamonds) and a sideband-cooled (circles) \mg ion. The thermal average over many Rabi frequency components results in a flat line for an excitation time beyond $15 \mu$s in case of the Doppler-cooled ion, whereas for the sideband-cooled ion a single-frequency oscillation is observed. The average error resulting from the dark and bright state distribution fit is on the order of $\pm 3\%$ for each point and is omitted in the graph for clarity. The solid line corresponds to an exponentially decaying sinusoidal, yielding a Rabi frequency of $\Omega = 2\pi \times 40.9(1)$\,kHz and a decay rate of $\Gamma = 4.2(4) \times 10^3\,s^{-1}$.}
\label{fig:gsc_carr_flops}       
\end{figure}

The fact that the ion is predominantly in a single state after sideband cooling also manifests itself in a significant difference in the observed Rabi oscillations on the Raman carrier transitions. Although carrier transitions do not change the motional quantum number, their Rabi frequency depends on the motional quantum number, as shown in \rfig\ref{fig:rabi_freq}. In case the ion is Doppler-cooled, an average over many oscillations with different Rabi frequencies is given, whereas if the ion occupies a Fock state, an oscillation with a single frequency is expected. This behavior can be observed in the Raman-stimulated Rabi oscillations on the carrier transition, as depicted in \rfig\ref{fig:gsc_carr_flops}. We attribute the decay of the oscillation for the sideband-cooled ion to several effects: dephasing due to residual $n>0$ population, magnetic field and intensity fluctuations and loss of coherence due to off-resonant excitation from the Raman beams. This last effect has been measured to increase (decrease) the ion excitation by $\sim\!0.02\%/\mu$s ($\sim\!0.06\%/\mu$s) for an ion in the \brightstate (\darkstate\hspace{-0.1cm}) state to first order. Taking this into account, the analysis of the ground state population yields $\bar{n}_{\textrm{\tiny ax}} = 0.02 \pm 0.01$.

It should be noted that off-resonant excitation could be significantly reduced by operating the system at an optical carrier detuning of $18.4$\,GHz and using the \second order EOM sidebands for cooling or by employing an EOM with a higher resonance frequency.

\section{Conclusion}
\label{conclusion}

In summary, we have demonstrated ground-state cooling of a single trapped \mg ion employing a frequency-quadrupled commercial fiber laser system. An EOM in the laser system allows for fast switching between a resonant cooling/repumping configuration and an off-resonant Raman laser configuration, respectively. This is a cost-effective simplification to other solid-state laser setups for Mg$^+$ ions \cite{Friedenauer:06}. The performance of the laser system was demonstrated in a pulsed ground-state cooling experiment with results comparable to the best achieved with more complex laser systems.

A further simplification of the optical setup is possible if internal state detection is not necessary and only ground state cooling is required. In this case, RSB pulses can be applied between two Zeeman components of the ground state of an ion without hyperfine structure, such as $^{24}$Mg$^+$, eliminating the need to bridge the hyperfine structure with AOMs. The demonstrated system can be used for a large variety of experiments involving ground-state cooling and coherent manipulation of ions.

\begin{acknowledgement}
We would like to thank Rainer Blatt for generous loan of equipment and support, L. An der Lan and B. Brandst\"atter for their help during the initial stages of the experiment and Uwe Sterr for providing the iodine spectroscopy cell. We further acknowledge the technical support of S. Haslwanter, A. Sch\"onherr, H. Jordan, M. Kluibensch\"adl, A. Wander, P.-C. Carstens, G. Hendl and S. Klitzing. We gratefully acknowledge financial support by the Austrian START program of the Austrian Ministry of Education and Science, the Cluster of Excellence QUEST, Hannover, and the Physikalisch-Technische Bundesanstalt, Braunschweig. The final publication is available at springerlink.com.
\end{acknowledgement}


\end{document}